\documentclass[aip,12pt]{revtex4-1}
\usepackage{graphicx}
\usepackage{amsmath}
\usepackage{placeins}
\begin{document}
\title{Room-temperature spin-orbit torque in NiMnSb}
\author{C.~Ciccarelli}
\thanks{These authors have contributed equally to the work.}
\affiliation{Cavendish Laboratory, University of Cambridge, CB3 0HE, United Kingdom}
\author{L.~Anderson}
\thanks{These authors have contributed equally to the work.}
\affiliation{Cavendish Laboratory, University of Cambridge, CB3 0HE, United Kingdom}
\author{V. Tshitoyan}
\affiliation{Cavendish Laboratory, University of Cambridge, CB3 0HE, United Kingdom}
\author{A.~J. Ferguson}
\thanks{e-mail: ajf1006@cam.ac.uk}
\affiliation{Cavendish Laboratory, University of Cambridge, CB3 0HE, United Kingdom}
\author{F.~Gerhard}
\affiliation{Physikalisches Institut (EP3), Universit\"{a}t W\"{u}rzburg, Am Hubland, D-97074 W\"{u}rzburg, Germany}
\author{C.~Gould}
\affiliation{Physikalisches Institut (EP3), Universit\"{a}t W\"{u}rzburg, Am Hubland, D-97074 W\"{u}rzburg, Germany}
\author{L.~W. Molenkamp}
\affiliation{Physikalisches Institut (EP3), Universit\"{a}t W\"{u}rzburg, Am Hubland, D-97074 W\"{u}rzburg, Germany}
\author{J.~Gayles}
\affiliation{Institut f\"ur Physik, Johannes Gutenberg Universit\"at Mainz, 55128 Mainz, Germany}
\author{J.~\v{Z}elezn\'y}
\author{L.~\v{S}mejkal}
\affiliation{Institute of Physics, Academy of Sciences of the Czech Republic, Cukrovarnicka 10, 162 00 Praha 6, Czech Republic}
\affiliation{Faculty of Mathematics and Physics, Charles University in Prague,
Ke Karlovu 3, 121 16 Prague 2, Czech Republic}
\author{Z.~Yuan}
\affiliation{Institut f\"ur Physik, Johannes Gutenberg Universit\"at Mainz, 55128 Mainz, Germany}
\author{J.~Sinova}
\affiliation{Institut f\"ur Physik, Johannes Gutenberg Universit\"at Mainz, 55128 Mainz, Germany}
\affiliation{Institute of Physics, Academy of Sciences of the Czech Republic, Cukrovarnicka 10, 162 00 Praha 6, Czech Republic}
\author{F.~Freimuth}
\affiliation{Peter Gr\"{u}nberg Institut and Institute for Advanced Simulation,
Forschungszentrum J\"{u}lich and JARA, 52425 J\"{u}lich, Germany}
\author{T.~Jungwirth}
\affiliation{Institute of Physics, Academy of Sciences of the Czech Republic, Cukrovarnicka 10, 162 00 Praha 6, Czech Republic}
\affiliation{School of Physics and Astronomy, University of Nottingham, Nottingham NG7 2RD, United Kingdom}
\date{\today}
\maketitle

\textbf{Materials that crystalize in diamond-related lattices, with Si and GaAs as  their prime examples, are at the foundation of modern electronics. Simultaneoulsy, the two atomic sites in the unit cell  of these crystals form inversion partners which gives rise to  relativistic non-equilibrium spin phenomena highly relevant for magnetic memories and other spintronic devices.   When the inversion-partner sites are occupied by the same atomic species, electrical current can generate local spin polarization with the same magnitude and opposite sign on the two inversion-partner sites. In CuMnAs, which shares this specific crystal symmetry of the Si lattice, the effect led to the demonstration of electrical switching in an antiferromagnetic memory at room temperature.\cite{Wadley2015}  When the inversion-partner sites are occupied by different atoms, a non-zero global spin-polarization is generated by the applied current which can switch a ferromagnet, as reported at low temperatures in the diluted magnetic semiconductor (Ga,Mn)As.\cite{Chernyshov2009} Here we demonstrate the effect of the global current-induced spin polarization in a counterpart crystal-symmetry material NiMnSb which is a member of the broad family of magnetic Heusler compounds. It is an ordered high-temperature ferromagnetic metal whose other favorable characteristics include  high spin-polarization  and low damping of magnetization dynamics. Our experiments are performed on strained single-crystal epilayers of NiMnSb grown on InGaAs. By performing all-electrical ferromagnetic resonance measurements in microbars patterned along different crystal axes we detect room-temperature spin-orbit torques generated by effective fields of the Dresselhaus symmetry. The measured magnitude and symmetry  of the current-induced torques are consistent with our relativistic density-functional theory calculations.
}

NiMnSb is a magnetic Heusler compound which in bulk has a Curie temperature of 730~K, a nearly 100\% spin polarization, a very low Gilbert damping constant $\sim 10^{-3}$, and which can have tuneable magnetic anisotropies when grown in thin films of varying thicknesses and stoichiometry.\cite{Gerhard2014} For these characteristics, NiMnSb has been utilized in earlier spintronics studies of magneto-transport and spin dynamics effects based on spin angular momentum transfer between carriers and magnetization in magnetic-multilayer devices.\cite{Hordequin98,thesis_Riegler}

Recently, the focus of  spintronic research has shifted from these essentially non-relativistic angular momentum transfer phenomena to effects which exploit the relativistic transfer between spin and the linear momentum of the electron. The spin-orbit coupling term, which originates is in the Dirac equation,  can generate spin polarizations and corresponding effective fields acting on the magnetization in single-layer magnetic systems\cite{Bernevig2005a,Chernyshov2009} or in multilayers\cite{Manchon2008,Miron2011b,Liu2012} when driven out of equilibrium by the applied current.\cite{Sinova2014} The multilayers, typically comprising an interface of a transition metal ferromagnet and a strongly spin-orbit coupled paramagnet, have attracted most of the attention to date. However, the single-layer magnets we focus on in this work have their own merits in the research of current induced spin polarization phenomena. The effects in these systems originate from the symmetries of unit cells of bulk crystals and the understanding and utility of current induced spin polarizations is, therefore, more robust against unintentional disorder  and structural imperfections than  the spin-orbit coupling phenomena generated within a few atomic planes near the ferromagnet/paramagnet interface.

In Figs.~1a,b we illustrate examples of the relativistic non-equilibrium spin polarizations that occur in the family of diamond-related lattices for the case when the two inversion partner lattice sites of the unit cell are occupied by the same or by different atomic species. The inverse spin-galvanic effect (also called the Edelstein effect),\cite{Silov2004,Kato2004b,Ganichev2004b,Wunderlich2004,Wunderlich2005,Ivchenko2008,Belkov2008} responsible for these polarization effects, requires the spin-orbit coupling to be combined with inversion asymmetries in the crystal structure.\cite{Belkov2008} Each of the two atomic lattice sites in the unit cell of the crystals shown in Fig.~1a,b have inversion asymmetric local environment\cite{Zhang2014} which allows for the generation of the local current-induced spin polarizations.\cite{Zelezny2014}

\begin{figure}[h!]
\centering
\includegraphics[trim=0cm 0cm 3cm 6cm,width=1\textwidth]{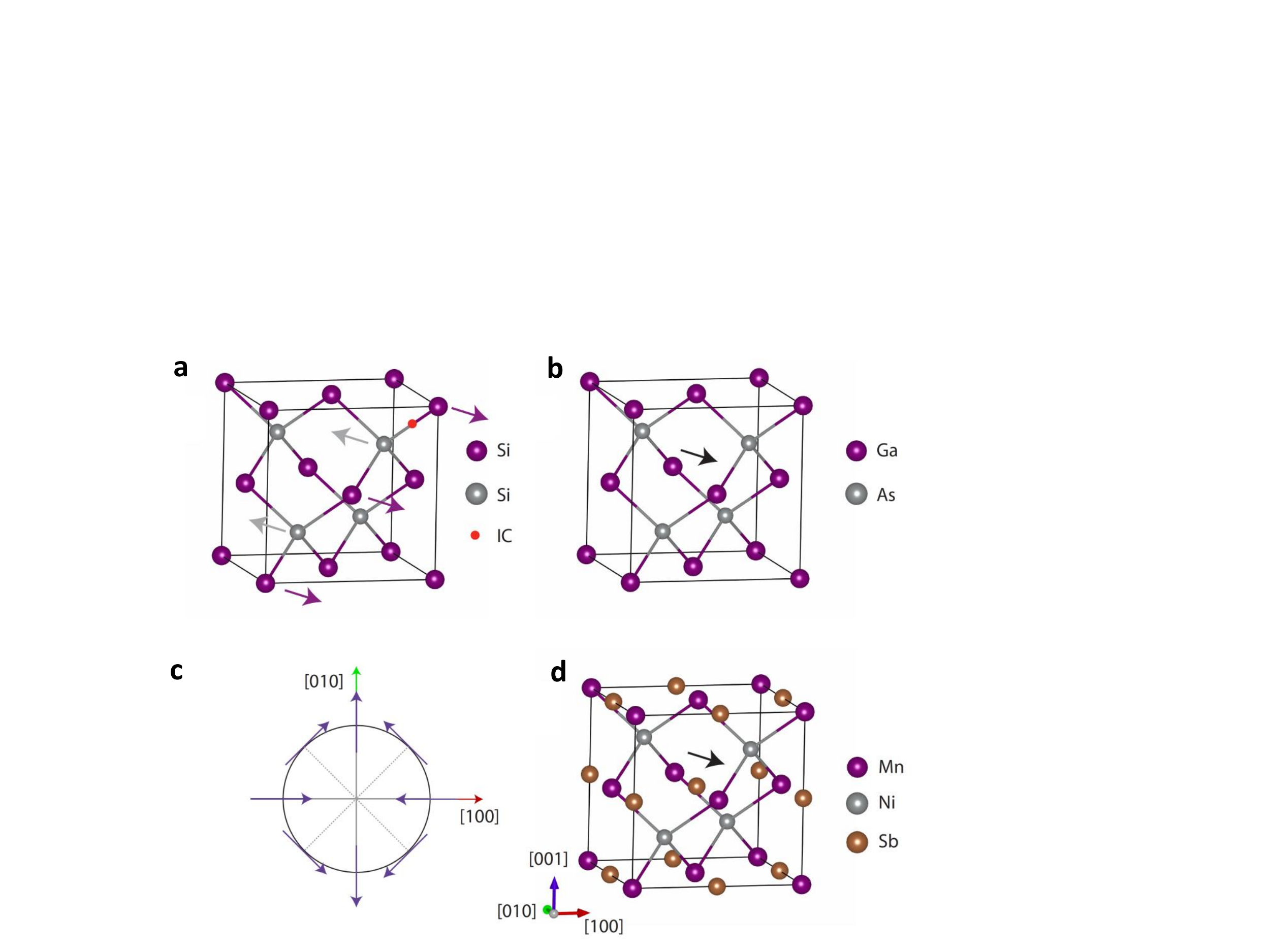}
\parbox{14cm}{\caption{Relativistic non-equilibrium spin polarisations in diamond-related lattices. (a) Crystal unit cell of Si. The red dot indicates the inversion center (IC) of the unit cell. In this case, the inversion partners are occupied by the same atomic species. By symmetry, the current-induced spin polarisation (arrows) has the same magnitude and opposite sign on the two inversion partner sites, which results in zero global spin polarisation. (b) Crystal unit cell of GaAs. In this case the inversion partners are occupied by different atomic species, which makes the unit cell globally non-centrosymmetric and results in a non-zero net spin polarisation. (c) Dresselhaus symmetry of the global spin polarisation with respect to the crystal direction of the applied current in tetragonally distorted diamond-related cubic lattices such as strained GaAs. (d) Crystal unit cell of half-Heusler NiMnSb. The symmetry of the cell is analogous to that of GaAs shown in (b).}
\label{fig1}}
\end{figure}

By symmetry, the global polarization vanishes when integrated over the entire unit  cell if the two sites are occupied by the same atom since the unit cell has an inversion center (highlighted by a red dot in Fig.~1a).  Only local polarizations of the same magnitude and opposite sign at the two inversion-partner sites remain non-zero in this case. When the sites are occupied by different atoms, the unit cell is globally inversion asymmetric allowing for the generation of a net global spin polarization with a non-zero integral value over the unit cell (see Fig.~1b). For completeness we point out that in these diamond-related cubic lattices an additional symmetry lowering has to be introduced to allow for the above  non-equilibrium polarization effects. For example, in thin films with tetragonal distortion due to a substrate lattice-matching strain, both the local and global spin polarizations acquire a Dresselhaus symmetry with respect to the crystal direction of the applied current, as illustrated in Fig.~1c. In Tab.~3 of the Supplementary information we summarize these conclusions based on the rigorous analysis of crystal symmetries of diamond-related lattices, as well as of all other crystals belonging to the 21 point groups with broken inversion symmetry.

The staggered local polarizations of the type shown in Fig.~1a can generate staggered fields that couple strongly to a N\'eel magnetic order, as predicted\cite{Zelezny2014} and subsequently demonstrated\cite{Wadley2015}  in room-temperature current induced switching experiments in a CuMnAs antiferromagnet.  The global polarization illustrated in Fig.~1b can be used to electrically switch ferromagnetic moments which, however, has so far been observed only at low-temperatures and in a strongly disordered random-moment ferromagnet (Ga,Mn)As. The aim of our work is to demonstrate the applicability of the effect in ferromagnetic crystals beyond the singular and academic example of (Ga,Mn)As. Guided by the general symmetry analysis of Tab.~3 of the Supplementary information, we choose NiMnSb. The symmetry of the NiMnSb lattice shown in Fig.~1d is analogous to the zinc-blende lattice of (Ga,Mn)As, however, NiMnSb is an ordered ferromagnet with high Curie temperature. This implies the possibility to generate global non-equilibrium carrier-polarizations and to use them for the electrical manipulation  of the ferromagnetic moments in NiMnSb at room temperature.

Our samples consist of an insulating InP substrate, 200~nm of an In$_{0.53}$Ga$_{0.47}$As buffer lattice-matched to the substrate, and 37~nm of a fully strained NiMnSb film capped with 5~nm of MgO. The material was grown in a multi-chamber molecular-beam-epitaxy system, allowing for  transfer between different chambers under ultra-high vacuum.\cite{Gerhard2014}  The crystal quality of the epilayers was confirmed by high resolution X-ray diffraction and RHEED measurements.

In order to deduce the vector of the current-induced effective field in NiMnSb, we perform current driven ferromagnetic resonance (FMR) measurements on two-terminal micro-bars (4~$\mu$m$\times$40~$\mu$m) patterned by electron beam lithography and ion milling. Similarly to investigations of spin torques in spin-valves,\cite{Tulapurkar2005} in ferromagnet/paramagnet bilayers,\cite{Liu2011} and in our previous experiments  in (Ga,Mn)As\cite{Fang2011,Kurebayashi2014} we excite magnetisation dynamics by passing a microwave frequency current of density $J$ through the micro-bar, as  shown in Fig.~2a. During precession of the magnetization, frequency mixing between the alternating current and the oscillating anisotropic magnetoresistance (AMR) produces a rectified component of the voltage, $V_{dc}$.

\begin{figure}[h!]
\centering
\includegraphics[trim=0cm 0cm 3cm 6cm,width=1\textwidth]{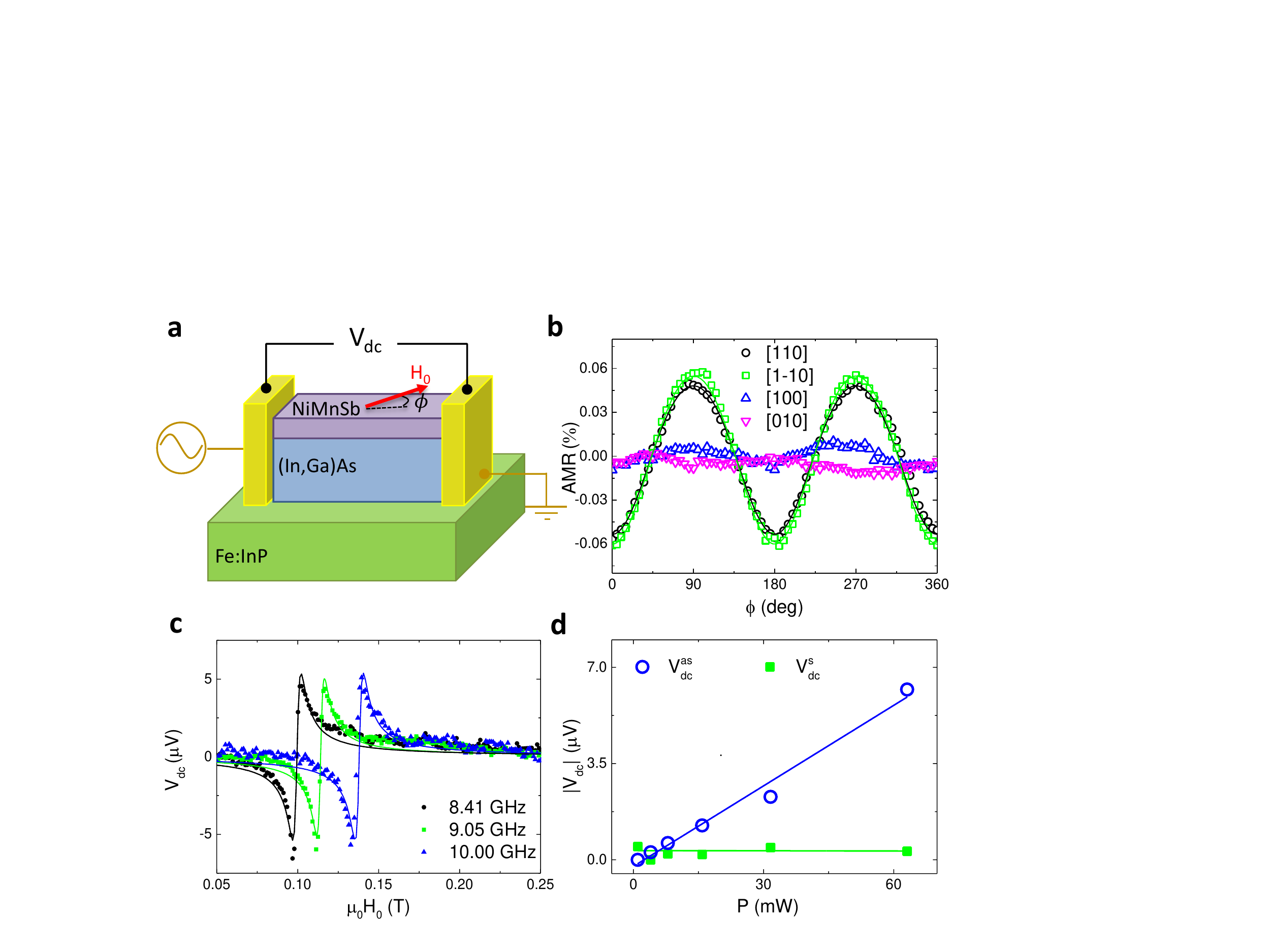}
\parbox{14cm}{\caption{Spin-orbit FMR experiment. (a) Schematic of the sample and measuring set-up. 4 $\times$ 40 $\mu m ^{2}$ bars are patterned from (In,Ga)As/NiMnSb epilayers on an insulating Fe-doped InP substrate. A microwave current is passed in the bar and excites spin-orbit FMR. By measuring the longitudinal dc voltage we are able to deduce the amplitude of precession, thus the magnitude of the spin-orbit driving field. The angle $\phi$ is the angle between the current flow and the external field around which the magnetisation precesses. (b) AMR measured in the [110], [1$\bar{1}0$], [100], and [010] oriented bars. (c) The rectified voltage showing FMR for different frequencies of the microwave current in a bar along the [110] direction. The Lorentzians are well fitted by an antisymmetric line-shape (continuous line) at all frequencies. (d) Power dependence of the symmetric and antisymmetric components of the rectified voltage.}
\label{fig2}}
\end{figure}

First, in Fig.~2b, we show static measurements of the AMR in bars patterned along [110], [1$\bar{1}0$], [100], and [010] crystal directions. We define $AMR\equiv\frac{R(\phi)-\bar{R}}{\overline{R}}$ where $R$ is the longitudinal resistance, $\phi$ is the angle between the current and the equilibrium magnetization set by a saturating magnetic field, and $\bar{R}$ is the longitudinal resistance averaged over $\phi$. All measurements presented in this work were performed at room temperature using an electromagnet to generate the magnetic field and a rotating stage to set $\phi$.

For the [110] and [1$\bar{1}0$] oriented bars, $AMR\approx C\cos(2\phi)$ where $C\approx 0.06\%$.  In the FMR experiments, the amplitude with which the resistance oscillates at resonance due to AMR is therefore given by $dAMR\approx-2C\sin(2\phi) d\phi$, where $d\phi$ is the small precession amplitude directly proportional to the effective current-induced field. By studying the angle dependence of $V_{dc}$ we can therefore find the symmetry of the effective field driving the precession.\cite{Fang2011,Kurebayashi2014}

We note that only  bars oriented along the [110] and [1$\bar{1}0$] axes were used in the FMR experiments. The AMR in [100] and [010] oriented bars nearly vanishes, as seen in Fig.~2b. In the Supplementary information we explain, based on symmetry analysis of the AMR measurements and on {\em ab initio} AMR calculations, that this is a consequence of the cancellation of non-crystalline and crystalline AMR terms in the NiMnSb film. Nevertheless, FMR measurements in the [110] and [1$\bar{1}0$] oriented bars are sufficient for inferring the magnitude and symmetry of the driving spin-orbit fields.

Fig.~2c shows the rectified voltage, $V_{dc}$, for a bar along the [110] crystal direction as the external magnetic field is swept through the FMR condition. The resonance is well fitted by an antisymmetric Lorentzian. The independence of the line-shape on the frequency of the current indicates that the phase between the current passed in the bar and the current-induced driving field is fixed and is not affected by reactive components of the circuit.\cite{Harder2011} This is a necessary requirement to carry the line-shape analysis outlined in Refs.~\onlinecite{Fang2011,Kurebayashi2014}.
The amplitude of the resonance is proportional to the incident microwave power (see Fig.~2d) implying that the driving field is linear in current density, as is typical for most mechanisms of current induced torques, including the Oersted field-torque and the spin torques.

Fig.~3a shows the rectified component of the voltage measured for a bar along the [110] crystal axis as the external magnetic field is swept at different in-plane directions with respect to the bar. In these measurements the frequency of the current is fixed at $\omega=2\pi\cdot9$~GHz, with a source power of 20 dBm. A resonance is clearly visible at fields above the saturation field of 30~mT. The resonance field depends on the anisotropy of the bar and its angle dependence can be fitted with the modified Kittel's formula deduced from the susceptibility matrix (for details see Supplementary Information). For the [110] and [1$\bar{1}$0] oriented bars we obtain from the fitting $\mu_0H_{2\perp}=638\pm3$~mT and $640\pm3$~mT,  $\mu_0H_{2\parallel}=21.4\pm0.3$~mT and $0.1\pm0.4$~mT, and $\mu_0H_{4\parallel}=11.1\pm0.5$~mT and $7.2\pm0.5$~mT, respectively. Here $\mu_0H_{2\perp}$ is the leading out-of-plane uniaxial anisotropy field due to the lattice-matching growth strain, and $\mu_0H_{2\parallel}$ and $\mu_0H_{4\parallel}$ are the weaker in-plane uniaxial and cubic anisotropy fields. Note that the in-plane anisotropies are different in the two microbars since their values in the unpatterned film are modified differently by partial ($\sim10$\%) strain relaxation in the 4~$\mu$m wide bars. The Gilbert damping constant inferred from  the dependence of the FMR line-width  on frequency is $(1.8\pm 0.10)\times 10^{-3}$, in agreement with previous reports \cite{Gerhard2014} (see Supplementary Information).

\begin{figure}[h!]
\centering
\includegraphics[trim=0cm 2cm 3cm 8cm,width=1\textwidth]{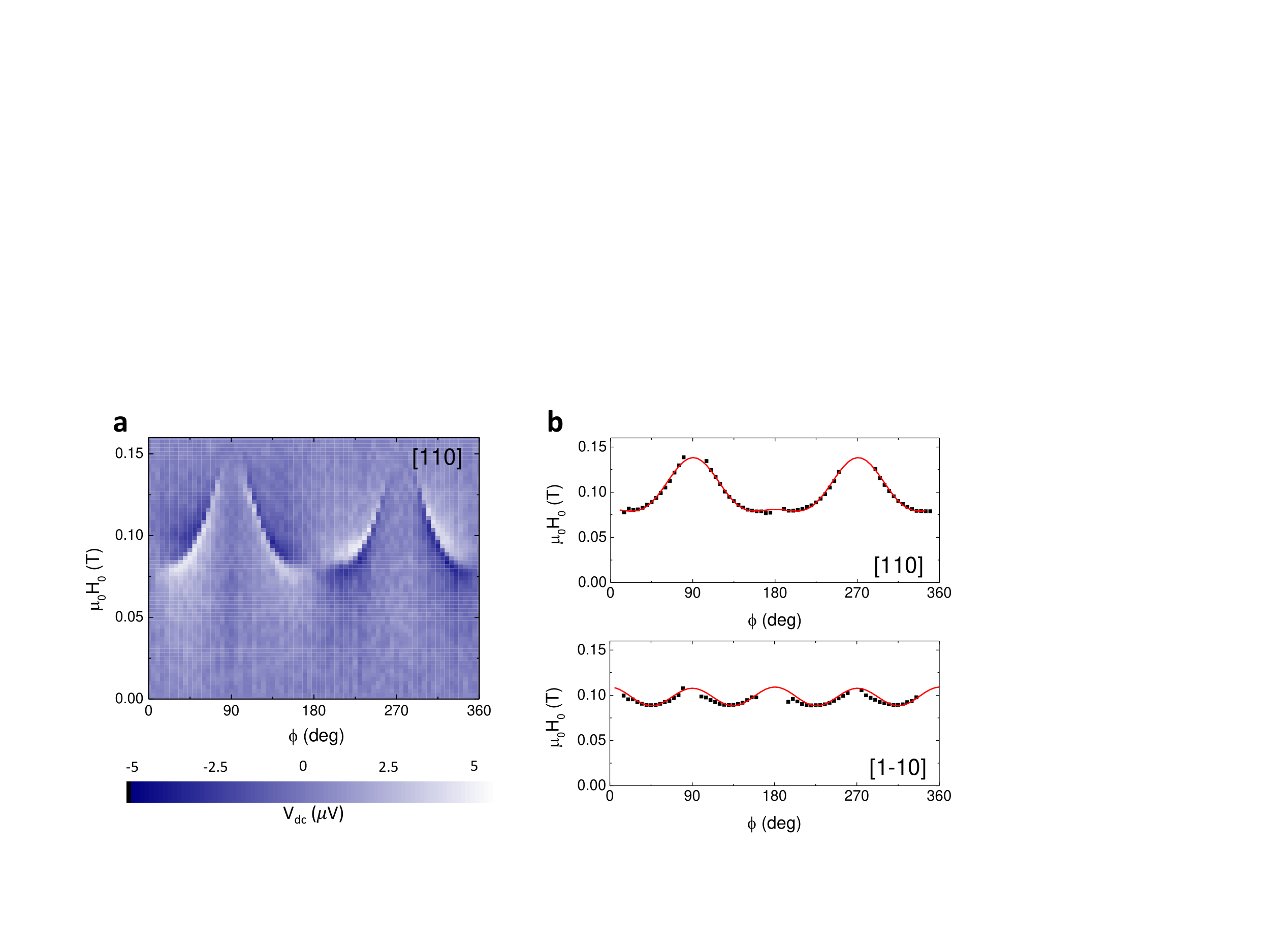}
\parbox{14cm}{\caption{Angle dependence of the resonance field. (a) Rectified voltage measured for a bar oriented along the [110] crystal direction as a function of the external field's amplitude and direction. (b)-(c) Resonance field measured for a bar along the [110] crystal axis (b) and along the [1-10] crystal axis (c) as a function of the angle $\phi$. The continuous line represents the fitting with the modified Kittel's formula.}
\label{fig3}}
\end{figure}

In Fig.~4a we plot the amplitude of the resonance with respect to the angle $\phi$ for the [110] and [1$\bar{1}$0] oriented bars. In both cases the amplitude exhibits a $\sin(2\phi)\cos(\phi)$ dependence, indicating maximum amplitude of precession, thus maximum torque, with the external magnetic field parallel to the bar. By fitting these graphs with the expression of $V_{dc}$ deduced from the susceptibility matrix,\cite{Fang2011,Kurebayashi2014} we determine the value of the current-induced fields to be $h_{[110]}=(340\pm20)$~$\mu T$ and $h_{[1\bar{1}0]}=(-550\pm50)$~$\mu T$ per $J=10^7$~Acm$^{-2}$. We were able to deduce the current density in the bar by performing heating calibration measurements as detailed in the Supplementary Information.

The current-induced fields measured for the two bars have opposite signs, excluding the Oersted field as a possible driving mechanism of precession and confirming their crystal origin. Fig.~4b shows the complete polar plot of the current-induced field in our bars, constructed by using the measured values $h_{[110]}$ (red arrow) and  $h_{[1\bar{1}0]}$ (black arrow)  as an orthogonal basis. Measurements on multiple sets of samples patterned along the [110] and [1$\bar{1}$0] crystal directions provided reproducible evidence for a room-temperature field-like torque driven by an effective field with the leading Dresselhaus symmetry, as illustrated in Fig.~4b.

\begin{figure}[h!]
\centering
\includegraphics[trim=0cm 4cm 3cm 8cm,width=1\textwidth]{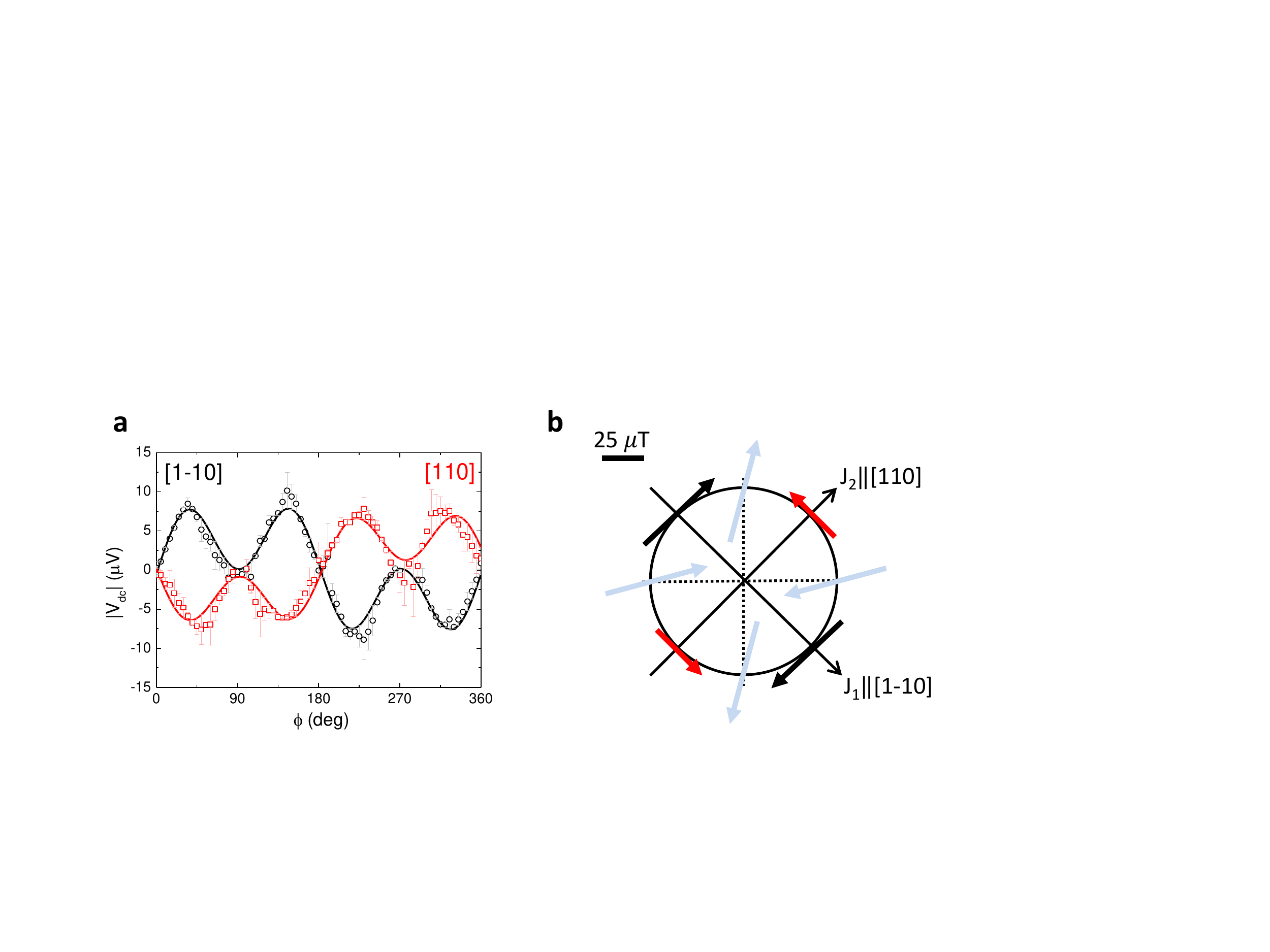}
\parbox{14cm}{\caption{Spin-orbit field components.(a) Antisymmetric component of the resonance in the rectified voltage for a bar along the [110] crystal axis (red squares) and a bar along the [1-10] crystal axis (black circles). From the fitting with the general expression found in [14,15], represented by the continuous line, we obtain magnitude and direction of the spin-orbit field for current flowing along the [110] direction, $\textbf{h}_{[110]}$, and the [1-10] direction, $\textbf{h}_{[1-10]}$. (b) Polar plot illustrating the direction of the spin-orbit field for current flowing along different crystal directions of NiMnSb. The fields represented by the red and black arrows were experimentally obtained. The fields represented by the light blue arrows were deduced by using $\textbf{h}_{[110]}$ and $\textbf{h}_{[1-10]}$ as an orthogonal basis.}
\label{fig4}}
\end{figure}

We now proceed to the theoretical analysis of our experimental data. The allowed symmetries of the current induced spin-polarizations for the 21 non-centrosymmetric crystallographic point groups are summarized in Tab.~3 of the Supplementary Information. The cubic half-Heusler lattice of NiMnSb shares the -43m point-group symmetry with the cubic diamond (zinc-blende) lattices. Under a tetragonal distortion, corresponding e.g. to our lattice-matching growth strain, the symmetry of  these crystals reduces to -42m. As seen from Tab.~3 in the Supplementary Information, the current-induced polarization is non-zero in the -42m point group and has the Dresselhaus symmetry, consistent with our experiments.

To obtain a theoretical estimate of the magnitude of the effective field which drives the spin-orbit torque in our NiMnSb film we performed relativistic density-functional-theory (DFT) calculations using two complementary approaches. In one method \cite{Freimuth2014a} we base our calculations on the full-potential linearized augmented plane-wave DFT code FLEUR for the description of the electronic structure. The spin-orbit torque is then calculated using the Kubo formalism for the linear response of the torque operator to the electric field. Effects of disorder are approximated by a constant quasiparticle broadening. In another method,\cite{Xia2006,Liu2011c} the electronic structure is determined in the DFT framework based on the tight-binding linear muffin-tin orbital approach. In the torque calculation, a scattering region is constructed  with the desired disorder and connected to semi-infinite perfectly crystalline leads.  The non-equilibrium spin polarization that is carried by conduction electrons is obtained from the explicit scattering wave functions using the wave function matching scheme. Thermal disorder is treated in the frozen phonon approximation. More details on these {\em ab initio} methods are given in the Supplementary Information.

In both calculations we considered the symmetry-lowering mechanism of the bulk cubic lattice of NiMnSb due to the substrate-matching growth strain. The resulting current-induced fields are of the Dresselhaus symmetry, in agreement with the crystallographic point group analysis and with experiment. The magnitude of the current-induced field obtained by the two {\em ab initio} methods is 89 and 111~$\mu$T per $J=10^7$~Acm$^{-2}$, respectively.  Without any adjustable parameter, the theoretical results agree on the order of magnitude level with our experiments.

To conclude, we have presented  a general framework, based on the complete set of crystallographic point groups, for identifying the potential occurrence and symmetry of current induced spin-orbit fields in globally or locally non-centrosymmetric crystals. While the zinc-blende lattice of, e.g., GaAs has served as an example of the globally non-centrosymmetric crystals, the diamond lattice of, e.g., Si and Ge has been introduced as an example of crystals with locally non-centrosymmetric lattice sites. We have pointed out that, generally,  the former type of inversion-asymmetric lattices is favorable for inducing efficient field-like current induced torques in ferromagnets while the latter type allows for inducing efficient field-like current induced torques in antiferromagnets. Based on these general arguments we have searched for a ferromagnetic crystal in which the spin-orbit torques can be detected at room temperature. We have identified NiMnSb for its globally non-centrosymmetric crystal structure reminiscent of GaAs, for its high ferromagnetic Curie temperature, and a range of other favorable structural and magnetic properties.  By performing room-temperature electrical FMR measurements in NiMnSb epilayers, we have detected current induced effective fields of the theoretically expected Dresselhaus symmetry and of a magnitude consistent with microscopic {\em ab initio} calculations. This has implications beyond designing spintronic devices in NiMnSb. Our results can guide the search for other favorable room temperature magnets that can be switched by the internal current-induced spin-orbit fields and do not require for the switching external magnetic fields or auxiliary magnetic polarizers embedded in complex magnetic multilayer structures.

\protect\newpage

\section{Methods}
\subsection{Materials}
The 37-nm-thick NiMnSb epilayer (room temperature conductivity of $2\times10^{4}$ $\Omega^{-1}$ $cm^{-1}$) was grown on 200 nm $In_{0.53}Ga_{0.47}As$ buffer layer (room temperature conductivity of 0.1 $\Omega^{-1}$ $cm^{-1}$) and Fe:InP insulating substrate and capped with 5 nm of MgO. The vertical lattice constant of 5.951 ${\AA}$ indicates a slightly Ni-rich composition.

\subsection{Experimental procedure}
A pulse-modulated (at 880 Hz) microwave signal was launched onto a printed circuit board patterned with a coplanar waveguide and then injected into the sample via a bond wire. The rectified voltage, generated at FMR, was separated from the microwave circuit by using a bias tee, amplified with a voltage amplifier and then detected with a lock-in amplifier. All measurements were performed at room temperature. A rotating stage allowed setting the orientation of the bar with respect to the fixed in-plane magnetic field generated by an electro-magnet.

\bibliography{Refs}


\section{Acknowledgments}  C.C. acknowledges support from  a  Junior research  fellowship  at  Gonville  and  Caius  College. L.A. acknowledges  support  from  the James B. Reynolds Scholarship at Dartmouth College. A.J.F.  acknowledges  support  from  a  Hitachi  research  fellowship and a EU ERC Consolidator Grant No.648613. F.G. acknowledges financial support from the University of W$\ddot{u}$rzburg's program \textit{Equal opportunities for women in research and teaching}. J.G. and J.S. acknowledge support from the the Alexander von Humboldt Foundation. L\v{S} acknowledges support from the Grant Agency of the Charles University, No. 280815 and access to computing and storage facilities owned by parties and projects contributing to the National Grid Infrastructure MetaCentrum, provided under the programme "Projects of Large Infrastructure for Research, Development, and Innovations" (LM2010005). J.Z. and F.F. gratefully acknowledge computing time on the supercomputers JUQUEEN and JUROPA at Juelich Supercomputing Centre. T.J. acknowledges support from EU ERC Advanced Grant No. 268066, from the Ministry of Education of the Czech Republic Grant No. LM2011026, and from the Grant Agency of the Czech Republic Grant no. 14-37427.

\section{Author contributions}  Theory and data modelling were performed by J.G., J.Z., L.S., Z.Y., J.S., F.F. and T.J. Materials were prepared by F.G. and C.G. Sample preparation was performed by C.C. Experiments and data analysis were carried out by C.C., L.A., V.T. and A.F. The manuscript was written by T.J.and C.C., project planning was performed by A.J.F., L.W.M., J.S. and T.J. All  authors discussed the results and commented on the paper.

\section{Competing Financial Interests} The authors declare no competing financial interests.

\end{document}